\RequirePackage{ifpdf}
\ifpdf 
\documentclass[pdftex]{sigma}
\else
\documentclass{sigma}
\fi

\numberwithin{equation}{section}

\def\bbt{\bibitem}
\def\d{\partial}
\def\rbr{\rbrack}
\def\lbr{\lbrack}

\def\ov{\over }
\def\tld{\tilde}

\def\sgm{\sigma}

\def\al{\alpha}
\def\bet{\beta}
\def\gm{\gamma}
\def\Gm{\Gamma}
\def\im{\imath}

\def\lm{\lambda}
\def\Lm{\Lambda}
\def\Om{\Omega}
\def\om{\omega}
\def\et{\eta}
\def\tt{\theta}

\def\ups{\upsilon}
\def\dlt{\delta}
\def\Dl{\Delta}
\def\kp{\kappa}

\def\bh{{\bf h}}
\def\bp{{\bf p}}

\def\bt{{\bf t}}

\def\bv{{\bf v}}
\def\bd{{\bf d}}

\def\rng{\rangle}
\def\lng{\langle}

\begin{document}

\allowdisplaybreaks

\renewcommand{\PaperNumber}{025}

\FirstPageHeading

\renewcommand{\thefootnote}{$\star$}

\ShortArticleName{Free Field Construction of D-Branes in Rational
Models of CFT and Gepner Models}

\ArticleName{Free Field Construction of D-Branes\\ in Rational
Models of CFT and Gepner Models\footnote{This paper is a
contribution to the Proceedings of the Seventh International
Conference ``Symmetry in Nonlinear Mathematical Physics'' (June
24--30, 2007, Kyiv, Ukraine). The full collection is available at
\href{http://www.emis.de/journals/SIGMA/symmetry2007.html}{http://www.emis.de/journals/SIGMA/symmetry2007.html}}}

\Author{Sergei  E. PARKHOMENKO}

\AuthorNameForHeading{S.E. Parkhomenko}

\Address{Landau Institute for Theoretical Physics Chernogolovka,
Russia} \Email{\href{mailto:spark@itp.ac.ru}{spark@itp.ac.ru}}

\ArticleDates{Received October 30, 2007, in f\/inal form February
14, 2008; Published online February 23, 2008}

\Abstract{This is a review article of my recent papers on free
f\/ield construction of D-branes in $N=2$ superconformal minimal
models and Gepner models.}

\Keywords{strings; D-branes; conformal f\/ield theory; free
f\/ield construction; minimal mo\-dels; Gepner models}

\Classification{17B65; 17B69; 17B81; 81R10; 81T30; 81T40; 81T60}

\section{Introduction}

\looseness=1
 The role of D-branes~\cite{Pol1} in the description of certain
nonperturbative degrees of freedom of strings is by now well
established and the study of their dynamics has led to many new
insights into String Theory. Much of this study was done in the
large volume limit where geometric technique provides reliable
information. The extrapolation into the stringy regime usually
requires boundary conformal f\/ield theory (CFT) methods. In this
approach D-brane conf\/igurations are given by conformally
invariant boundary states or boundary conditions. However a
complete microscopic description of these conf\/igurations are
well understood only for the case of f\/lat and toric backgrounds
where the CFT on the world sheet is a theory of free f\/ields.

 The class of rational CFT's gives the interesting examples of curved
string backgrounds where the construction of the boundary states
can be given in principle. One of the important examples is $N=2$
supersymmetric minimal models which are building blocks of the
Gepner models~\cite{Gep} of superstring compactif\/ication on
toric Calabi--Yau manifolds. Thus, the problem of construction of
D-branes in various rational models of CFT and in $N=2$
supersymmetric minimal models, in particular, is an important and
interesting.

 However the explicit construction of boundary states in these models is more complicated
problem as against the case of f\/lat or toric backgrounds.
Although the structure of Hilbert space in rational models of CFT
is naturally governed by the chiral algebra of symmetries the
construction of the space of states is very nontrivial problem
because it consists of highly degenerate representations of chiral
algebra. This special feature of rational models hampers
considerably the description of irreducible chiral algebra modules
in terms of chiral algebra generators as the inf\/inite number of
singular vectors and submodules is generated by the action of
chiral algebra on the highest weight vector. In other words, the
modules freely generated by the chiral algebra are highly
reducible. They contain inf\/inite number of submodules which must
be properly  factored out to get the irreducible representations
comprising the Hilbert space. The structure of singular vectors
and submodules as well as the way of factorization can be
described by resolutions of the irreducible
modules~\cite{FeFu,FeM,FeFr,BFeld,FeSST,FeS}.

 The free f\/ield representations and corresponding resolutions are of special
interest in rational models of CFT. They are not only code the
submodules structure but allow also to construct explicitly the
f\/ields of the model via the intertwining operators acting
between the irreducible modules~\cite{FeFu,FeFr,BFeld}. This
property is very important because it provides a way to f\/ind the
correlation functions in an explicit form. In other words, free
f\/ield representation allows to f\/ind the solutions of the
corresponding dif\/ferential equations in an explicit
form~\cite{DF,Dot,BFeld,FeFr} which is a~nontrivial problem.

 In the rational models of CFT with boundary the singular vectors problem appears again when
we try to construct the Ishibashi and boundary states because we
need to introduce an orthonormal basis for each irreducible
representation. One can avoid this problem if we use the
resolutions instead of the irreducible representations. Hence it
is natural to apply the free f\/ield approach to the models with
boundary. This problem has been solved in~\cite{SP,SP1}, for the
case of $SL(2)$ WZNW models and $N=2$ supersymmetric minimal
models. The idea is to change the irreducible modules by the Fock
modules resolutions and use these resolutions for Ishibashi and
boundary states construction.

Notice that the construction is quite general and can be applied
to any rational model of CFT with known free f\/ield realization.
In addition, the free f\/ield construction can be extended for the
case of Gepner models~\cite{SP2,SP3}, where the
Recknagel--Schomerus boundary states as well as Recknagel
permutation branes have been constructed in an explicit form by
free f\/ields. Because of boundary states in Gepner models, are
def\/ined by  purely algebraic construction, the question of their
geometric interpretation is nontrivial and interesting. I hope
that free f\/ield approach might appears to be natural and
ef\/f\/icient for the description of D-brane geometry in these
models.

\looseness=1
 This is a review article of the papers~\cite{SP1} and~\cite{SP2}. Section \ref{sec2} is devoted
to the free f\/ield construction of boundary states in $N=2$
superconformal minimal models. We start with the preliminary
material on $N=2$ minimal models in Subsection~\ref{sec2.1}. In
Subsection \ref{sec2.2} we consider free f\/ield realization of
the right-moving and left-moving $N=2$ Virasoro superalgebra
currents and introduce the Fock modules, where the superalgebras
are naturally acting. Then, we construct the Fock modules
Ishibashi states satisfying A and B-type boundary conditions. In
the next Subsection we use free f\/ield resolutions (so called
butterf\/ly resolutions) of irreducible $N=2$ Virasoro algebra
modules to represent free f\/ield construction of Ishibashi states
for corresponding irreducible modules. This is the main part of
the construction. The problem here is to cancel the contributions
from the huge number of redundant (non-physical) closed string
states coming from the singular vectors. Similar to the bulk case,
the non-physical states decoupling condition is equivalent to BRST
invariance of the Ishibashi state, with respect to the sum of BRST
charges of butterf\/ly resolutions in the left and right-moving
sectors. Then, free f\/ield construction of boundary states is
given by Cardy prescription to the Ishibashi states. More recent
paper~\cite{IWat} should be mentioned in this respect. In this
paper the free f\/ield approach to $SL(2)$ WZNW models has also
been considered but the BRST invariance condition which is of
crucial importance has not been properly taken into account. In
Subsection~\ref{sec2.4} we consider the closed string geometry of
D-branes in $N=2$ minimal model using the free f\/ield
realization.

\looseness=1 Section~\ref{sec3} deals with the construction by
free f\/ields the D-branes in Gepner models. The construction is a
straightforward generalization of the $N=2$ minimal model case. It
is brief\/ly discussed in Subsection \ref{sec3.1}. Then, we
consider in Subsection~\ref{sec3.2} free f\/ield geometry of
D-branes in closed and open string sectors. It is motivated by the
conjecture that geometry of D-branes at string scales has
natural description in terms of the free f\/ields.

\newpage

\section[Free field construction of D-branes in $N=2$ supersymmetric minimal models]{Free f\/ield construction of D-branes\\ in $\boldsymbol{N=2}$ supersymmetric minimal models}\label{sec2}

\subsection[The symmetry algebra, Hilbert space and boundary conditions in
$N=2$ supersymmetric  minimal models]{The symmetry algebra,
Hilbert space and boundary conditions\\ in $\boldsymbol{N=2}$
supersymmetric  minimal models}\label{sec2.1}

The algebra of symmetries is given by holomorphic (or left-moving) and
antiholomorphic (or right-moving) copies of $N=2$ Virasoro
superalgebra
\begin{gather}
T(z)=\sum_{n}L[n]z^{-n-2}, \qquad J(z)=\sum_{n}J[n]z^{-n-1},
\nonumber\\
G^{+}(z)=\sum_{r}G^{+}[r]z^{-r-3/2}, \qquad
G^{-}(z)=\sum_{r}G^{-}[r]z^{-r-3/2},
\nonumber\\
\bar{T}(\bar{z})=\sum_{n}\bar{L}[n]\bar{z}^{-n-2}, \qquad
\bar{J}(\bar{z})=\sum_{n}\bar{J}[n]\bar{z}^{-n-1},
\nonumber\\
\bar{G}^{+}(\bar{z})=\sum_{r}\bar{G}^{+}[r]\bar{z}^{-r-3/2},
\qquad \bar{G}^{-}(\bar{z})=\sum_{r}\bar{G}^{-}[r]\bar{z}^{-r-3/2}
\label{1.Vir}
\end{gather}
with the central charge
\begin{gather*}
c=3\left(1-\tfrac{2}{\mu}\right), \qquad \mu=2,3,\dots
\end{gather*}
and the following commutation relations
\begin{gather}
\{G^{+}[r],G^{-}[s]\}=\left(r^{2}-\tfrac{1}{4}\right)\tfrac{c}{6}\dlt_{r+s,0}+
\tfrac{(r-s)}{2}J[r+s]+L[r+s], \nonumber\\
\lbr J[n],G^{\pm}[r]\rbr=\pm G^{\pm}[n+r], \qquad
\lbr L[n],G^{\pm}[r]\rbr=\left(\tfrac{n}{2}-r\right)G^{\pm}[n+r], \nonumber\\
\lbr L[n],J[m]\rbr=-mJ[n+m], \qquad
\lbr J[n],J[m]\rbr=n\tfrac{c}{3}\dlt_{n+m,0}, \nonumber\\
\lbr
L[n],L[m]\rbr=n(n^{2}-1)\tfrac{c}{12}\dlt_{n+m,0}+(n-m)L[n+m].
\label{1.com}
\end{gather}
 As usual, the holomorphic and antiholomorphic fermionic currents in NS sector are expanded into
half-integer modes and hence $r$, $s$ are half-integer numbers in
(\ref{1.Vir}), (\ref{1.com}). While they are expanded into integer
modes in $R$ sector, where $r$, $s$ are integer numbers.

 Hilbert space of the model in NS-NS sector is a direct sum of tensor products of irreducible $N=2$ Virasoro
superalgebra representations
\begin{gather*}
\Om =\oplus_{h=0}^{\mu-2}\oplus_{j=h}^{-h}M_{h,j}\otimes
\bar{M}_{h,j}.
\end{gather*}
The irreducible modules $M_{h,j}$ are generated by the $N=2$
Virasoro superalgebra generators from the highest vectors
$|h,j\rng$ determined by the following annihilation conditions
\begin{gather}
L[n]|h,j\rng=0, \qquad J[n]|h,j\rng=0, \quad n>0,
\nonumber\\
G^{\pm}[r]|h,j\rng=0, \quad r>-\tfrac 12,\qquad
L[0]|h,j\rng={h(h+2)-j^{2}\ov 4\mu}, \qquad  J[0]|h,j\rng={j\ov
\mu}|h,j\rng. \label{1.hvect}
\end{gather}
When $j=h$ ($j=-h$) we have a so called chiral (anti-chiral)
module generated from the chiral (anti-chiral) highest vector
$|h,h\rng$ ($|h,-h\rng$) satisfying additional relation
$G^{+}[-1/2]|h,h\rng=0$ ($G^{-}[-1/2]|h,-h\rng=0$).

 Notice that module $M_{h,j}$ is not freely generated from
the highest vector. One can show~\cite{FeSST} that there is an
inf\/inite number of submodules in the module freely generated
from the highest vector $|h,j\rng$. For example the vector which
is called a singular vector
\begin{gather}
|E^{+}\rng=G^{+}\left[-\tfrac 12-m\right]\cdots G^{+}\left[-\tfrac
32\right]|h,h\rng. \label{1.singvect}
\end{gather}
satisfies the conditions similar to (\ref{1.hvect}) and hence
generates a submodule. Therefore, one has to factor out this one
as well as the other submodules to get the irreducible
representation $M_{h,h}$ from freely generated module. This
problem is not trivial because of the number of submodules is
inf\/inite and they are intersect to each other~\cite{FeSST,FeS}.

 The $N=2$ Virasoro algebra has the following set of automorphisms
which is known as spectral f\/low~\cite{SS}
\begin{gather}
G^{\pm}[r]\rightarrow U^{-t}G^{\pm}[r]U^{t}=G^{\pm}[r\pm t], \nonumber\\
L[n]\rightarrow
U^{-t}L[n]U^{t}=L[n]+tJ[n]+t^{2}\tfrac{c}{6}\dlt_{n,0},\qquad
J[n]\rightarrow U^{-t}J[n]U^{t}=J[n]+t \tfrac{c}{3}\dlt_{n,0},
\label{1.flow}
\end{gather}
where $t\in Z$. Allowing  $t$ in (\ref{1.flow}) to be
half-integer, we obtain the isomorphism between the Hilbert space
in NS and R sectors. Moreover, one can use the
observation~\cite{FeS} that all irreducible modules can be
obtained from the chiral modules $M_{h,j=h}$, $h=0,\dots,\mu-2$ by
the spectral f\/low action $U^{-t}$, $t=h,h-1,\dots,1$.
Equivalently, one can restrict the set of chiral modules by the
range $h=0,\dots,[{\mu\ov2}]-1$ (here, $[{\mu\ov 2}]$ denotes the
integer part of $\mu\ov 2$) and extend the spectral f\/low action
by $t=\mu-1,\dots,1$. Thus, the set of irreducible modules can be
labeled also by the set $\{(h,t)|h=0,\dots,[{\mu\ov2}]-1,\
t=\mu-1,\dots,0 \}$ (when $\mu$ is even and $h=[{\mu\ov2}]-1$ the
spectral f\/low orbit becomes shorter:
$t=[{\mu\ov2}]-1,\dots,1$~\cite{FeSST}).

There are two types of boundary states preserving $N=2$
super-Virasoro algebra usually called $B$-type
\begin{gather}
(L[n]-\bar{L}[-n])|B\rng\rng=0, \qquad
(J[n]+\bar{J}[-n])|B\rng\rng=0, \nonumber\\
(G^{+}[r]+\im \et \bar{G}^{+}[-r])|B\rng\rng=0, \qquad
(G^{-}[r]+\im \et \bar{G}^{-}[-r])|B\rng\rng=0, \qquad
|B\rng\rng\in \Om \label{1.BD}
\end{gather}
and $A$-type states
\begin{gather}
(L[n]-\bar{L}[-n])|A\rng\rng=0, \qquad
(J[n]-\bar{J}[-n])|A\rng\rng=0,
\nonumber\\
(G^{+}[r]+\im \et \bar{G}^{-}[-r])|A\rng\rng=0, \qquad
(G^{-}[r]+\im \et \bar{G}^{+}[-r])|A\rng\rng=0,\qquad
|A\rng\rng\in \Om, \label{1.AD}
\end{gather}
where $\im=\sqrt{-1}$ and $\et=\pm 1$.

\subsection[Free field realization and Ishibashi states in Fock modules]{Free f\/ield realization and Ishibashi states in Fock modules}\label{sec2.2}

Our f\/irst problem is to f\/ind a solution of the equations
(\ref{1.BD}), (\ref{1.AD}) in the product of irreducible $N=2$
Virasoro superalgebra modules. In other words we are going to
construct the Ishibashi state. However, the explicit construction
of Ishibashi state is quite complicated problem not only because
the $N=2$ Virasoro superalgebra is noncommutative but also because
of the problem of singular vectors occurs. Instead, we use free
f\/ield realization of $N=2$ minimal models and construct f\/irst
the Ishibashi states for Fock modules which is much easy to do.

Free f\/ield realization in the holomorphic sector is given by
free bosonic f\/ields $X(z)$, $X^{*}(z)$ and free fermionic
f\/ields $\psi(z)$, $\psi^{*}(z)$,
\begin{gather*}
X(z)=x+X_{0}\ln(z)+\sum_{n\neq 0}{X[n]\ov n}z^{-n}, \qquad
X^{*}(z)=x^{*}+X^{*}_{0}\ln(z)+\sum_{n\neq 0}{X^{*}[n]\ov
n}z^{-n},
\nonumber\\
\psi(z)=\sum_{r\in 1/2+Z}\psi[r]z^{-{1\ov 2}-r}, \qquad
\psi^{*}(z)=\sum_{r\in 1/2+Z}\psi^{*}[r]z^{-{1\ov 2}-r},
\\
\lbr X^{*}[n],X[m]\rbr=n\dlt_{n+m,0},\! \qquad \lbr
X[0],x^{*}\rbr=1, \!\qquad \lbr X^{*}[0],x\rbr=1,\! \qquad
\lbr\psi^{*}[r],\psi[s]\rbr_{+}=\dlt_{r+s,0}.\nonumber
\end{gather*}

The currents of $N=2$ super-Virasoro algebra are given by
\begin{gather*}
G^{+}(z)=\psi^{*}(z)\d X(z) -\tfrac{1}{\mu} \d \psi^{*}(z), \qquad
G^{-}(z)=\psi(z) \d X^{*}(z)-\d \psi(z),
\nonumber\\
J(z)=\psi^{*}(z)\psi(z)+\tfrac{1}{\mu}\d X^{*}(z)-\d X(z),
\nonumber\\
T(z)=\d X(z)\d X^{*}(z)+ \tfrac{1}{2}\left(\d^{2}
X(z)+\tfrac{1}{\mu}\d^{2} X^{*}(z)\right)+ \tfrac{1}{2}(\d
\psi^{*}(z)\psi(z)-\psi^{*}(z)\d \psi(z)).
\end{gather*}
It is clear that $N=2$ Virasoro superalgebra is acting in the Fock
module $F_{p,p^{*}}$ which is gene\-rated~by creation operators of
the f\/ields $X(z)$, $X^{*}(z)$, $\psi(z)$, $\psi^{*}(z)$ from the
vacuum vector $|p,p^{*}\rng\in F_{p,p^{*}}$. It is determined by
\begin{gather*}
\psi[r]|p,p^{*}\rng=\psi^{*}[r]|p,p^{*}\rng=0, \quad r\geq
\tfrac{1}{2}, \qquad X[n]|p,p^{*}\rng=X^{*}[n]|p,p^{*}\rng=0,
\quad
n\geq 1, \nonumber\\
X_{0}|p,p^{*}\rng=p|p,p^{*}\rng, \qquad
X^{*}_{0}|p,p^{*}\rng=p^{*}|p,p^{*}\rng.
\end{gather*}
The vector $|p,p^{*}\rng$ satisfies the conditions (\ref{1.hvect})
where
\begin{gather*}
h=p^{*}+\mu p, \qquad j=p^{*}-\mu p.
\end{gather*}

 Analogously, the Fock module $\bar{F}_{\bar{p},\bar{p}^{*}}$,
which is generated from vacuum $|\bar{p},\bar{p}^{*}\rng$ by the
creation operators of the antiholomorphic f\/ields $\bar{X}$,
$\bar{X}^{*}$, $\bar{\psi}$, $\bar{\psi}^{*}$, is the
representation for $N=2$ Virasoro algebra in the right-moving
sector.

 In the tensor product of Fock modules we
are going to construct Ishibashi state
\begin{gather*}
|p,p^{*},\bar{p},\bar{p}^{*}\rng\rng\in
F_{p,p^{*}}\otimes\bar{F}_{\bar{p},\bar{p}^{*}}
\end{gather*}
which fulf\/ills the relations (\ref{1.BD}) for example. One can
show that (\ref{1.BD}) follows from
\begin{gather}
(\psi^{*}[r]- \im
\et\bar{\psi}^{*}[-r])|p,p^{*},\bar{p},\bar{p}^{*},B\rng\rng=0,
\qquad (\psi[r]- \im
\et\bar{\psi}[-r])|p,p^{*},\bar{p},\bar{p}^{*},B\rng\rng=0,
\nonumber\\
\left(X[n]+\bar{X}[-n]+{\dlt_{n,0}\ov \mu}\right)
|p,p^{*},\bar{p},\bar{p}^{*},B\rng\rng=0,
\nonumber\\
(X^{*}[n]+\bar{X}^{*}[-n]+\dlt_{n,0})
|p,p^{*},\bar{p},\bar{p}^{*},B\rng\rng=0,
\nonumber\\
\bar{p}=-p-\tfrac{1}{\mu}, \qquad \bar{p}^{*}=-p^{*}-1.
\label{1.BX}
\end{gather}

The solution is given by a standard expression
\begin{gather}
|p,p^{*},B\rng\rng=
\prod_{n=1}\exp\left(-\tfrac{1}{n}X^{*}[-n]\bar{X}[-n]\right)
\exp\left(-\tfrac{1}{n}X[-n]\bar{X}^{*}[-n]\right)
\nonumber\\
\phantom{|p,p^{*},B\rng\rng=}{}\times\prod_{r=1/2}\exp(\im\et\psi^{*}[-r]\bar{\psi}[-r])
\exp(\im\et\psi[-r]\bar{\psi}^{*}[-r]))|p,p^{*},-p-\tfrac{1}{\mu},-p^{*}-1\rng.\!\!\!
\label{1.Bish}
\end{gather}

 The $A$-type Ishibashi states can be found analogously.
The relations (\ref{1.AD}) follow from
\begin{gather}
(\psi^{*}[r]- \im
\et\mu\bar{\psi}[-r])|p,p^{*},\bar{p},\bar{p}^{*},\et,A\rng\rng=0,
\qquad \left(\psi[r]- \im
\tfrac{\et}{\mu}\bar{\psi}^{*}[-r]\right)|p,p^{*},\bar{p},\bar{p}^{*},\et,A\rng\rng=0,
\nonumber\\
(\mu X[n]+\bar{X}^{*}[-n]+\dlt_{n,0})
|p,p^{*},\bar{p},\bar{p}^{*},\et,A\rng\rng=0, \nonumber\\
(X^{*}[n]+\mu\bar{X}[-n]+\dlt_{n,0})
|p,p^{*},\bar{p},\bar{p}^{*},\et,A\rng\rng=0,
\nonumber\\
\bar{p}=-{1+p^{*}\ov \mu}, \qquad \bar{p}^{*}=-\mu -1.
\label{1.AX}
\end{gather}
Hence the $A$-type Fock space Ishibashi state (in NS sector) is
given by
\begin{gather*}
|p,p^{*},\et,A\rng\rng=
\prod_{n=1}\exp\left(-\tfrac{1}{n}\left(\mu X[-n]\bar{X}[-n]+\tfrac{1}{\mu}X^{*}[-n]\bar{X}^{*}[-n]\right)\right) \nonumber\\
\phantom{|p,p^{*},\et,A\rng\rng=}{}\times\!\prod_{r=1/2}\!
\exp\left(\im\et\!\left(\tfrac{1}{\mu}\psi^{*}[-r]\bar{\psi}^{*}[-r]+
\mu\psi[-r]\bar{\psi}[-r]\right)\!\right)\!|p,p^{*},-{1+p^{*}\ov
\mu},-\mu p-1\rng.
\end{gather*}

\subsection[Ishibashi states in irreducible $N=2$ Virasoro superalgebra modules
and boundary states in $N=2$ minimal model]{Ishibashi states in
irreducible $\boldsymbol{N=2}$ Virasoro superalgebra modules\\ and
boundary states in $\boldsymbol{N=2}$ minimal model}
\label{sec2.3}

How to use the solution (\ref{1.Bish}) to build the Ishibashi
state for irreducible $N=2$ Virasoro superalgebra modules?

The Fock module $F_{p,p^{*}}$ is highly reducible representation.
It contains inf\/inite number of singular (and cosingular) vectors
generating $N=2$ Virasoro submodules similar to the
example~(\ref{1.singvect}). Therefore, the state~(\ref{1.Bish})
contains contributions from huge number of redundant states coming
from singular vectors.

 In other words, the overlap of this state with an
arbitrary closed string state which does not belong to the Hilbert
space of the $N=2$ minimal model is not zero in general
\begin{gather*}
\lng\lng p,p^{*},\et,B|v\otimes \bar{v}\rng\neq 0, \qquad
|v\otimes \bar{v}\rng \in
F_{p,p^{*}}\otimes\bar{F}_{\bar{p},\bar{p}^{*}}, \qquad |v\otimes
\bar{v}\rng\notin \Om.
\end{gather*}
If we want to build the Ishibashi state for irreducible
representation it is necessary to cancel contributions from
submodules.

Hence we have to investigate the structure of submodules in the
Fock modules and extract the irreducible representations. For the
module $F_{0,h}$ it is given by the following inf\/inite
complex~\cite{FeS} (which is known due to Feigin and Semikhatov as
butterf\/ly resolution)
\begin{gather}\!\!
\begin{array}{ccccccccccc}
&&\vdots &\vdots &&&&&&\\
&&\uparrow &\uparrow &&&&&&\\
\ldots &\leftarrow &F_{1,h+\mu} &\leftarrow
F_{0,h+\mu}&&&&&&\\
&&\uparrow &\uparrow &&&&&&\\
\ldots &\leftarrow &F_{1,h} &\leftarrow F_{0,h}&&&&&&\\
&&&&\nwarrow&&&&&\\
&&&&&F_{-1,h-\mu}&\leftarrow &F_{-2,h-\mu}&\leftarrow &\ldots\\
&&&&&\uparrow &&\uparrow&\\
&&&&&F_{-1,h-2\mu}&\leftarrow &F_{-2,h-2\mu}&\leftarrow &\ldots\\
&&&&&\uparrow &&\uparrow &&\\
&&&&&\vdots &&\vdots &&
\end{array} \label{1.but}
\end{gather}
The horizontal arrows in this diagram are given by the action of
screening charge
\begin{gather*}
Q^{+}=\oint dz\psi^{*}\exp{(X^{*})}
\end{gather*}
and vertical arrows are given by the action of screening charge
\begin{gather*}
Q^{-}=\oint dz \psi\exp{(\mu X)}.
\end{gather*}
The diagonal arrow at the middle of butterf\/ly resolution is
given by the action of $Q^{+}Q^{-}$. The ghost number $g$ of the
complex is increasing along the diagonal from the right to the
left and~$F_{0,h}$ module is ghost number zero subspace.

 The crucial property of the $Q^{\pm}$ operators is that
they commute to the Virasoro algebra generators and they are
nilpotent
\begin{gather*}
[L[n],Q^{\pm}]=[J[n],Q^{\pm}]=\cdots=0, \qquad (Q^{\pm})^{2}=0, \
Q^{+}Q^{-}+Q^{-}Q^{+}=0.
\end{gather*}
Hence the $Q^{\pm}$-images of the Fock modules are the $N=2$
Virasoro superalgebra submodules. In particular, $Q^{\pm}$-images
of the vacuum vectors $|p,p^{*}\rng$ are the singular vectors
like~(\ref{1.singvect}).

\begin{theorem}[\cite{FeS}]\label{theorem1}
Complex \eqref{1.but} is exact except at the $F_{0,h}$ module,
where the cohomology is given by the chiral module $M_{h,j=h}$.
\end{theorem}

What about the singular vector structure for the module
$\bar{F}_{-{1\ov \mu},-1-h}$ coming from the right-moving sector
of the state (\ref{1.Bish})? It is given by dual butterf\/ly
resolution~\cite{SP1}.
\begin{gather}\!\!
\begin{array}{cccc@{}c@{}cccccc}
&&\vdots &\vdots &&&&&&\\
&&\downarrow &\downarrow &&&&&&\\
\ldots &\rightarrow &\bar{F}_{-1-{1\ov\mu},-1-h-\mu} &\rightarrow
\bar{F}_{-{1\ov\mu},-1-h-\mu}&&&&&&\\
&&\downarrow &\downarrow &&&&&&\\
\ldots &\rightarrow &\bar{F}_{-1-{1\ov\mu},-1-h} &\rightarrow
\bar{F}_{-{1\ov\mu},-1-h}&&&&&&\\
&&&&\searrow&&&&&\\
&&&&&\bar{F}_{1-{1\ov\mu},-1-h+\mu}&\rightarrow
&\bar{F}_{2-{1\ov\mu},-1-h+\mu}
&\rightarrow &\ldots\\
&&&&&\downarrow &&\downarrow&\\
&&&&&\bar{F}_{1-{1\ov\mu},-1-h+2\mu}&\rightarrow
&\bar{F}_{2-{1\ov\mu},-1-h+2\mu}
&\rightarrow &\ldots\\
&&&&&\downarrow &&\downarrow &&\\
&&&&&\vdots &&\vdots &&
\end{array}\hspace{-10mm} \label{1.dualbut}
\end{gather}
The arrows on this diagram are given by the same operators as on
the diagram~(\ref{1.but}).

\begin{theorem}[\cite{SP1}]\label{theorem2}
Complex \eqref{1.dualbut} is exact except at the
$\bar{F}_{-{1\ov\mu},-1-h}$ module, where the cohomology is given
by the anti-chiral irreducible module $M_{h,j=-h}$.
\end{theorem}

To make the cancelation of redundant states from submodules we add
(with appropriate coef\/f\/icients) the Fock Ishibashi states
(\ref{1.Bish}) from the tensor products of modules forming the
butterf\/ly complex (\ref{1.but}) and its dual (\ref{1.dualbut}).
Thus, the free f\/ield construction of the Ishibashi state for the
irreducible representation
\begin{gather*}
|M_{h,j=h},\et,B\rng\rng\in M_{h,j=h}\otimes \bar{M}_{h,j=-h}
\end{gather*}
is given by the superposition
\begin{gather}
|M_{h,j=h},\et,B\rng\rng=
\sum_{n,m\geq0}c_{n,m}|n,m\mu+h,\et,B\rng\rng+
\sum_{n,m>0}c_{-n,-m}|-n,-m\mu+h,\et,B\rng\rng,\!\!\!
\label{1.MIB}
\end{gather}
where the coef\/f\/icients $c_{n,m}$ are f\/ixed by the BRST
invariance condition which is equivalent to the redundant states
cancelation.

 BRST invariance condition can be formulated as follows.
First of all, we form a tensor product of the butterf\/ly
complexes (\ref{1.but}) and (\ref{1.dualbut})
\begin{gather}
\cdots\rightarrow C_{h,h}^{-2}\rightarrow C_{h,h}^{-1}\rightarrow
C_{h,h}^{0}\rightarrow C_{h,h}^{+1}\rightarrow\cdots,
\label{1.bicomp}
\end{gather}
which is graded by the sum of the ghost numbers $g+\bar{g}$. The
dif\/ferential $D$ of the complex is def\/ined by the
dif\/ferentials $d$ and $\bar{d}$ of the complexes (\ref{1.but})
and (\ref{1.dualbut})
\begin{gather*}
D|v_{g}\otimes\bar{v}_{g'}\rng=|dv_{g}\otimes\bar{v}_{g'}\rng+
(-1)^{g}|v_{g}\otimes\bar{dv}_{g'}\rng,
\end{gather*}
where $|v_{g}\rng$ is an arbitrary vector from the complex
(\ref{1.but}) with the ghost number $g$, while $|\bar{v}_{g'}\rng$
is an arbitrary ghost number $g'$ vector from the complex
(\ref{1.dualbut}).

 It follows from the theorems above that the cohomology of the complex
(\ref{1.bicomp}) is centered at zero grading
\begin{gather*}
{\bf H}^{0}=M_{h,h=j}\otimes \bar{M}_{h,j=-h}, \qquad {\bf
H}^{g}=0, \qquad g\neq 0.
\end{gather*}

 The Ishibashi state we are looking for
can be considered as a linear functional on the Hilbert space of
$N=2$ superconformal minimal model. Therefore, the BRST invariance
condition can be written as follows:
\begin{gather}
D^{*}|M_{h,j=h},\et,B\rng\rng=0 \quad \Leftrightarrow \quad
\lng\lng M_{h,h},\et,B|D|v_{g}\otimes \bar{v}_{i-g}\rng=0, \qquad
|v_{g}\otimes \bar{v}_{i-g}\rng\in C_{h,h}^{i}.\!\! \label{1.D*}
\end{gather}
From this equation we f\/ind~\cite{SP1} that coef\/f\/icients of
the superposition $c_{n,m}$ are proportional to~$c_{0,0}$
according to the ghost numbers of the double complex
(\ref{1.bicomp})
\begin{gather}
c_{n,m}=\exp{\left(\im\et\tfrac{\pi}{2}(n+m)^{2}\right)}c_{0,0}.
\label{1.cnm}
\end{gather}
It gives explicit free f\/ield construction of Ishibashi state
$|M_{h,j=h}\rng\rng$.

The free f\/ield construction of the Ishibashi states for the
modules $M_{h,j}$, $j\ne h$ is given by the spectral f\/low action
$U^{t}$, where $t=0,\dots,h$
\begin{gather}
|M_{h,t},\et,B\rng\rng=U^{t}(\bar{U})^{-t}|M_{h},\et,B\rng\rng.
\label{1.Btish}
\end{gather}

 Analogously, the free f\/ield representation of $A$-type irreducible module Ishibashi
state is given by the following superposition of Fock Ishibashi
states
\begin{gather}
|M_{h,t},\et,A\rng\rng=U^{t}(\bar{U})^{t}|M_{h},\et,A\rng\rng,
\nonumber\\
|M_{h},\et,A\rng\rng=
\sum_{n,m\geq0}c_{n,m}|n,m\mu+h,\et,A\rng\rng+
\sum_{n,m>0}c_{-n,-m}|-n,-m\mu+h,\et,A\rng\rng, \label{1.MIA}
\end{gather}
where another normalization of the screening charges
$\bar{Q}^{\pm}$ has been used to cancel $\mu$ and ${1\ov \mu}$
factors for $c_{n,m}$ coef\/f\/icients.

 Now the free f\/ield construction of $A$ and $B$-type boundary states (in NS sector)
in $N=2$ minimal models is given by Cardy prescription~\cite{C}
\begin{gather}
|D_{h,t},\et,A\rng\rng=
\sum_{h',t'}D_{(h,t),(h',t')}|M_{h',t'},\et,A\rng\rng,
\nonumber\\
|D_{h,t},\et, B\rng\rng= \sum_{h'={\rm
even}}D_{(h,t),(h',t'=h'/2)}|M_{h',t=h'/2},\et, B\rng\rng,
\label{1.DCardy}
\end{gather}
where $D_{(h,t),(h',t')}$ are Cardy coef\/f\/icients determined by
$N=2$ minimal model modular matrix
\begin{gather*}
S_{(h,t),(h',t')}=
{1\ov\sqrt{2}\mu}\sin\left({\pi(h+1)(h'+1)\ov\mu}\right)\exp\left(\im\pi{(h-2t)(h'-2t')\ov\mu}\right).
\end{gather*}
Thus, the coef\/f\/icients $c_{0,0}$ in the superpositions
(\ref{1.MIA}) should be identif\/ied with the Cardy's
coef\/f\/icients.

 In conclusion of this section we notice that the free f\/ield
representations of boundary states are determined modulo
BRST-exact states satisfying $A$ or $B$-type boundary conditions.
We interpret this ambiguity in the free f\/ield representation as
a result of adding brane-antibrane pairs annihilating under the
tachyon condensation process~\cite{Sen}. Strictly speaking, the
BRST-exact states ambiguity is not the usual brane-antbrane pair
ambiguity and has to be considered in a generalized sense, because
BRST-exact states contain also the states with negative charges in
NS sector. In particular, there are the states in NS sector
oppositely charged with respect to graviton. Similar objects has
recently been discussed in~\cite{Ghost}. In this context the free
f\/ield representation (\ref{1.DCardy}) of boundary states can be
considered as the superposition of branes f\/lowing under the
(generalized) tachyon condensation to nontrivial boundary states
in minimal model.
  It is also important to note that spectral f\/low automorphisms (\ref{1.flow}) give
dif\/ferent free f\/ield representations of boundary states
because the corresponding butterf\/ly resolutions are not
invariant with respect to these automorphisms. However, their
cohomology is invariant. Hence, these dif\/ferent representations
have to be identif\/ied and the free f\/ield boundary state
construction should be considered in the sense of derived
category~\cite{DCat}.

\subsection[Free field geometry of D-branes in $N=2$ minimal models]{Free f\/ield geometry of D-branes in $\boldsymbol{N=2}$ minimal models}\label{sec2.4}

Introducing the new set of bosonic and fermionic oscillators
\begin{gather}
v[n]=\tfrac{\im}{\sqrt{2\mu}}(X^{*}[n]-\mu X[n]), \qquad
u[n]=\tfrac{1}{\sqrt{2\mu}}(X^{*}[n]+\mu X[n]),
\nonumber\\
\bar{v}[n]=\tfrac{\im}{\sqrt{2\mu}}(\bar{X}^{*}[n]-\mu
\bar{X}[n]), \qquad
\bar{u}[n]=-\tfrac{1}{\sqrt{2\mu}}(\bar{X}^{*}[n]+\mu \bar{X}[n]),
\nonumber\\
\sgm[r]=\tfrac{1}{\sqrt{2\mu}}(\psi^{*}[r]+\mu \psi[r]), \qquad
\gm[r]=\tfrac{\im}{\sqrt{2\mu}}(\psi^{*}[r]-\mu \psi[r]),
\nonumber\\
\bar{\sgm}[r]=-\tfrac{1}{\sqrt{2\mu}}(\bar{\psi}^{*}[r]+\mu
\bar{\psi}[r]), \qquad
\bar{\gm}[r]=\tfrac{\im}{\sqrt{2\mu}}(\bar{\psi}^{*}[r]-\mu
\bar{\psi}[r]). \label{1.tor}
\end{gather}
one can rewrite the $A$ and $B$-type boundary conditions
(\ref{1.AX}), (\ref{1.BX}) as
\begin{gather*}
(u[n]-\bar{u}[-n]+\sqrt{\tfrac{2}{\mu}}\dlt_{n,0})
|p,p^{*},\bar{p},\bar{p}^{*},\et,A\rng\rng=0, \qquad
(v[n]-\bar{v}[-n]) |p,p^{*},\bar{p},\bar{p}^{*},\et,A\rng\rng=0,
\nonumber\\
(\gm[r]+\im\et\bar{\gm}[-r])
|p,p^{*},\bar{p},\bar{p}^{*},\et,A\rng\rng=0, \qquad
(\sgm[r]-\im\et\bar{\sgm}[-r])
|p,p^{*},\bar{p},\bar{p}^{*},\et,A\rng\rng=0,
\nonumber\\
\left(u[n]-\bar{u}[-n]+\sqrt{\tfrac{2}{\mu}}\dlt_{n,0}\right)
|p,p^{*},\bar{p},\bar{p}^{*},\et,B\rng\rng=0, \qquad
(v[n]+\bar{v}[-n]) |p,p^{*},\bar{p},\bar{p}^{*},\et,B\rng\rng=0,
\nonumber\\
(\gm[r]+\im\et\bar{\gm}[-r])
|p,p^{*},\bar{p},\bar{p}^{*},\et,B\rng\rng=0, \qquad
(\sgm[r]-\im\et\bar{\sgm}[-r])
|p,p^{*},\bar{p},\bar{p}^{*},\et,B\rng\rng=0.
\end{gather*}
Thus, $A$-type states correspond to Dirichlet boundary condition
along the coordinates $u$, $v$ and $B$-type states correspond to
Dirichlet boundary condition along the coordinate $u$ and to
Neumann boundary condition along the coordinate $v$. Notice also
that we can view the coordinates $\exp(u)$, $v$ as polar
coordinates on the complex plane because of the currents $J(z)$,
$T(z)$ take the form
\begin{gather}
J(z)=\psi^{*}(z)\psi(z)-\im\sqrt{\tfrac{2}{\mu}}\d v(z), \qquad
T(z)=\tfrac{1}{2}(\d u^{2}+\d v^{2})-\tfrac{1}{\sqrt{2\mu}}\d^{2}u.
\label{1.JTtor}
\end{gather}
Hence, in these coordinates $A$-type states are the points on the
complex plane and $B$-type states correspond to 1-dimensional
circles around the origin. It is clear however, that we are free
to change arbitrary the sings in front of the right-moving
coordinates in (\ref{1.tor}) changing thereby the boundary
conditions. The def\/inition (\ref{1.tor}) stems from the
consistency with the expressions (\ref{1.MIB}), (\ref{1.Btish}),
(\ref{1.MIA}) and our wish to have a geometry close to geometry of
LG model. Indeed, due to the summation over $n+m$ in (\ref{1.MIB})
and (\ref{1.MIA}) it is natural to view the boundary conditions as
Dirichlet along the coordinate $u$, which is noncompact direction,
so that the winding states are not allowed in this direction by
topology. The summation over the $n-m$ is natural to consider as a
summation over the winding modes in $B$-type branes and summation
over the momenta in $A$-type branes. In this picture, $A$-type
boundary conditions correspond to D0-branes while $B$-type
boundary conditions correspond to D1-branes. It should be noted
once more that this geometric interpretation is quite artif\/icial
and it would be interesting to f\/ind a~CFT argument to f\/ix the
ambiguity of boundary conditions interpretation.

\section[Free field construction of D-branes in Gepner models]{Free f\/ield construction of D-branes in Gepner models}\label{sec3}

 In Gepner models of superstring compactif\/ication the string degrees of freedom on the compact
manifold are given by internal $N=2$ CFT which is direct product
of $N=2$ supersymmetric minimal models factored out by GSO
projection
\begin{gather*}
\Om\approx M_{1}\times M_{2}\times\cdots \times M_{I}/{\rm GSO},
\qquad c=\sum_{i=1}^{I}c_{i}=9, \qquad
c_{i}=3\left(1-\tfrac{2}{\mu_{i}}\right).
\end{gather*}

 The $N=2$ Virasoro superalgebra of the model
is given by the diagonal in the product of individual $N=2$
Virasoro superalgebras
\begin{gather*}
L[n]=(L_{(1)}+\cdots+L_{(I)})[n], \qquad
J[n]=(J_{(1)}+\cdots+J_{(I)})[n],\qquad \dots.
\end{gather*}
The GSO projection group in the internal sector is generated by
the operator
\begin{gather}
G_{\rm GSO}=\exp{(2\im \pi (J[0]+\bar{J}[0]))} \label{2.3}
\end{gather}
and hence the diagonal $N=2$ Virasoro algebra survives the
projection. Thus, we have purely algebraic construction of the
Hilbert space~\cite{Gep} (see also~\cite{EOTY,FSW}).

It is obvious that free f\/ield representation of D-branes in
Gepner models can be obtained from the free f\/ield construction
of D-branes in $N=2$ minimal models if we take into account the
GSO projection~\cite{SP2}.

\subsection[Free field representation of D-branes]{Free f\/ield representation of D-branes}\label{sec3.1}

 In the free f\/ield language we introduce in the left-moving sector the free bosonic
f\/ields $X_{i}(z)$, $X^{*}_{i}(z)$ and free fermionic f\/ields
$\psi_{i}(z)$, $\psi^{*}_{i}(z)$, $i=1,\dots,I$ as well as the
lattice of momenta
\begin{gather*}
\Pi =P\oplus P^{*},
\nonumber\\
P={\bp=\left({n_{1}\ov\mu_{1}},\dots,{n_{I}\ov\mu_{I}}\right)},
\quad n_{i}\in Z, \qquad P^{*}={\bp^{*}=(m_{1},\dots,m_{I})},
\quad m_{i}\in Z
\end{gather*}
and the set of Fock modules associated to this lattice
\begin{gather*}
F_{\bp,\bp^{*}}, \qquad (\bp,\bp^{*})\in \Pi.
\end{gather*}
The corresponding butterf\/ly resolution is given by the product
of butterf\/ly resolutions of individual minimal models. Hence the
dif\/ferential is given by the screening charges
\begin{gather*}
Q^{+}_{i}=\oint dz\psi^{*}_{i}\exp{(X^{*}_{i})}, \qquad
Q^{-}_{i}=\oint dz\psi_{i}\exp{(\mu_{i}X_{i})}.
\end{gather*}
The similar objects have to be introduced in the right-moving
sector.

 Thus, according to (\ref{1.MIA}) the free f\/ield construction of
$A$-type Ishibashi state in the product of minimal models is given
by
\begin{gather*}
|M_{\bh,\bt},\et,A\rng\rng=
\prod_{i}U_{i}^{-t_{i}}\bar{U}_{i}^{-t_{i}}
|M_{\bh},\et,A\rng\rng, \qquad |M_{\bh},\et,A\rng\rng=
\sum_{(\bp,\bp^{*})\in\Gm_{\bh}}c_{\bp,\bp^{*}}|\bp,\bp^{*},\et,A\rng\rng,
\end{gather*}
where $\Gm_{\bh}$ is the set of butterf\/ly resolution momenta and
\begin{gather*}
|\bp,\bp^{*},\et,A\rng\rng=
\prod_{n=1}\exp\left(-\tfrac{1}{n}\sum_{i}\left(\mu_{i}X_{i}[-n]\bar{X}_{i}[-n]+
\tfrac{1}{\mu_{i}}X^{*}_{i}[-n]\bar{X}^{*}_{i}[-n]\right)\right) \nonumber\\
{}\times\!
\prod_{r=1/2}\!\exp\left(\im\et\sum_{i}\left(\mu_{i}\psi_{i}[-r]\bar{\psi}_{i}[-r]+
\tfrac{1}{\mu_{i}}\psi^{*}_{i}[-r]\bar{\psi}^{*}_{i}[-r]\right)\right)
|\bp,\bp^{*},-\Om^{-1}\bp^{*}-\bd,-\Om\bp-\bd^{*}\rng,
\end{gather*}
where we have introduced the matrix $\Om_{ij}=\mu_{i}\dlt_{ij}$.
Now the construction of $A$-type boundary states is
straightforward, we use the Cardy prescription and the GSO
projection
\begin{gather*}
|{\bf\Lm},{\bf\lm},\et,A\rng\rng={\om\ov\kp^{2}}
\sum_{(\bh,\bt)\in \Dl}W^{\bh,\bt}_{{\bf\Lm},{\bf\lm}}
\sum_{m,n=0}^{\kp-1}\exp{(\im 2\pi
nJ_{0})}U^{m\bv}\bar{U}^{m\bv}|M_{\bh,\bt},\et,A\rng\rng,
\end{gather*}
where $W^{\bh,\bt}_{{\bf\Lm},{\bf\lm}}$ are Cardy's
coef\/f\/icients
\begin{gather*}
W^{\bh,\bt}_{{\bf\Lm},{\bf\lm}}=R^{\bh}_{{\bf\Lm}}
\exp{\left(\im\pi\sum_{i=1}{(\Lm_{i}-2\lm_{i})(h_{i}-2t_{i})\ov\mu_{i}}\right)},\qquad
R^{\bh}_{{\bf\Lm}}=\prod_{i=1}^{r}R^{h_{i}}_{\Lm_{i}}, \nonumber\\
R^{h_{i}}_{\Lm_{i}}={S_{\Lm_{i},h_{i}}\ov\sqrt{S_{0,h_{i}}}},
\qquad
S_{\Lm_{i},h_{i}}={\sqrt{2}\ov\mu_{i}}\sin\left({\pi(\Lm_{i}+1)(h_{i}+1)\ov\mu_{i}}\right),
\end{gather*}
$\Dl$ is the set of irreducible representations from the product
of minimal models and $\om$ is the normalization constant. The
summation over $n$ makes projection on the set of integer
$J[0]$-charges providing thereby GSO projection in the internal
sector of the superstring~\cite{FSW,ReS}, while the summation over
$m$ introduce spectral f\/low twisted sectors. The boundary states
are labelled by pair of vectors
$({\bf\Lm},{\bf\lm})=(\Lm_{1},\dots,\Lm_{I},\lm_{1},\dots,\lm_{I})\in
\Dl$.

The $B$-type boundary states are given by the similar
expression~\cite{SP2}. Thus, we obtain the construction of
Recknagel--Schomerus boundary states~\cite{ReS} in an explicit
form. The free f\/ield representation can be generalized also to
the case of permutation branes of Recknagel~\cite{Reper}. It was
given in~\cite{SP3}.

\subsection[Free field geometry of D-branes in Gepner models]{Free f\/ield geometry of D-branes in Gepner models}\label{sec3.2}

 The geometry of D-branes analysis in the closed string sector is given analogously to the
minimal models. Similar to (\ref{1.tor}) we introduce in the
left-moving sector the bosonic f\/ields $v_{i}(z)$, $u_{i}(z)$,
the fermionic f\/ields $\sgm_{i}(z)$, $\gm_{i}(z)$ and do the same
in the right-moving sector, introducing  $\bar{v}_{i}(\bar{z})$,
$\bar{u}_{i}(\bar{z})$, $\bar{\sgm}_{i}(\bar{z})$,
$\bar{\gm}_{i}(\bar{z})$, $i=1,\dots,I$. It is clear that in this
picture $A$-type boundary conditions in the Gepner model are
Dirichlet-ones for all coordinates, while $B$-type boundary
conditions are Dirichlet conditions for $u_{i}$ coordinates and
Neumann-ones for~$v_{i}$ coordinates. Hence, before the GSO
projection the $A$-type boundary state corresponds to D0-brane
in the f\/lat complex space~${\mathbb C}^{I}$ and $B$-type
boundary state describes a Lagrangian $I$-dimensional torus in
${\mathbb C}^{I}$, where $\exp{(u_{i})}$, and~$v_{i}$ are the
polar coordinates. Due to (\ref{1.tor}), (\ref{1.JTtor}) and
(\ref{2.3}) it is easy to see how the GSO group is acting in
${\mathbb C}^{I}$, it makes the orbifold ${\mathbb C}^{I}/{\rm
GSO}$. Hence the $A$ and $B$-type boundary states in Gepner models
are the D0 and D$I$-branes on the orbifold. The question which
seems to me important is to understand what is geometric meaning
of Cardy's coef\/f\/icients and how they code the moduli space of
D-branes in Gepner models.

 There is another aspect of the free f\/ield geometry of D-branes in Gepner models.
It appears in the open string sector of D-branes~\cite{SP2,SP3}.
In this case the space of states of the open strings between the
pair of D-branes can be described by one of the sectors, the
left-moving one for example. It is easy to calculate the
transition amplitude between the pair of boundary states. The
result comes from the BRST construction of Ishibashi states
(\ref{1.D*}), (\ref{1.cnm}) as well as representation of the
irreducible modules characters in the $N=2$ minimal model by an
alternating sum of the Fock modules characters from the
butterf\/ly resolution~\cite{FeS}:
\begin{gather}
Z^{A}_{({\bf\Lm}_{1},{\bf\lm}_{1})({\bf\Lm}_{2},{\bf\lm}_{2})}(q)\equiv
\lng\lng{\bf\Lm}_{1},{\bf\lm}_{1},\et,A|(-1)^{g}\tld{q}^{L_{0}-{c\ov24}}
|{\bf\Lm}_{2},{\bf\lm}_{2},\et,A\rng\rng
\nonumber\\
\qquad{}=\om^{2}\sum_{\bh,\bt}\prod_{i}(N_{\Lm_{1,i},\Lm_{2,i}}^{h_{i}}
\dlt^{(2\mu_{i})}(\Lm_{2,i}-2\lm_{2,i}-\Lm_{1,i}+2\lm_{1,i}+h_{i}-2t_{i})
\nonumber\\
\qquad{}+N_{\Lm_{1,i},\Lm_{2,i}}^{\mu_{i}-h_{i}-2}
\dlt^{(2\mu_{i})}(\Lm_{2,i}-2\lm_{2,i}-\Lm_{1,i}+2\lm_{1,i}+h_{i}-2t_{i}-\mu_{i}))
{\rm ch}_{\bh,\bt}(q), \label{2.amp}
\end{gather}
where $(-1)^{g}$ is the ghost number insertion according to BRST
condition (\ref{1.D*}), $q=\exp{(2\im\pi\tau)}$,
$\tld{q}=\exp{(-2\im{\pi\ov \tau})}$,
$N_{\Lm_{1,i},\Lm_{2,i}}^{h_{i}}$ are Verlinde algebra
coef\/f\/icients for the individual minimal model, and
\begin{gather*}
{\rm ch}_{\bh,\bt}(q,u)={1\ov\kp^{2}}\sum_{n,m=0}^{\kp-1} {\rm
Tr}_{M_{\bh,\bt}}(U^{n\bv}q^{(L_{0}-{c\ov24})}u^{J_{0}}\exp{(2\im\pi
mJ_{0})}U^{-n\bv})\\
\phantom{ch_{\bh,\bt}(q,u)}{} =
{1\ov\kp^{2}}\sum_{n,m=0}^{\kp-1}\chi_{\bh,\bt+n\bv}(\tau,\ups+m),
\end{gather*}
where $M_{\bh,\bt}=\prod_{i}M_{h_{i},t_{i}}$ is the product of
irreducible modules, $U^{n\bv}=\prod_{i}U_{i}^{n}$ is the product
of spectral f\/low operators and $\chi_{\bh,\bt}(\tau,\ups)$ is
the product of characters of minimal models. Hence, the transition
amplitude calculates the number of open string states between the
pair of D-branes weighted by $q^{L_{0}-{c\ov 24}}$.

 This space is given by the cohomology of the butterf\/ly
resolution (which is the product of the butterf\/ly resolutions
for individual minimal model).

The calculation of cohomology is given in two steps. At f\/irst
step we calculate the cohomology with respect to the
$\sum_{i}Q^{+}_{i}$-dif\/ferential. It is well known that
$\sum_{i}Q^{+}_{i}$-cohomology is generated by the set of
$bc\bet\gm$-system of f\/ields
\begin{gather}
a_{i}(z)=\exp{(X_{i}(z))}, \qquad a^{*}_{i}(z)=(\d
X^{*}_{i}-\psi_{i}\psi^{*}_{i})
\exp{(-X_{i}(z))}, \nonumber\\
\al_{i}(z)=\psi_{i}\exp{(X_{i}(z))}, \qquad
\al^{*}_{i}(z)=\psi^{*}_{i}\exp{(-X_{i}(z))}, \label{2.btgm}
\end{gather}
Their Fourier components commute with each other except the
following
\begin{gather*}
\lbr a^{*}[n]a[m]\rbr=\dlt_{n+m,0}, \qquad \lbr
\al^{*}[r],\al[s]\rbr_{+}=\dlt_{r+s,0}.
\end{gather*}
Hence, one can interpret these f\/ields as a string version of the
complex coordinates, the derivatives and dif\/ferentials on the
f\/lat complex space ${\mathbb C}^{I}$:
\begin{gather*}
a_{i}(z)\rightarrow  {\rm coordinates} \ a_{i} \ {\rm on} \
{\mathbb C}^{I}, \qquad a^{*}_{i}(z) \rightarrow \ {\rm
derivatives} \ {\d\ov \d a_{i}},
\nonumber\\
\al_{i}\rightarrow \ {\rm dif\/ferentials} \ da_{i}, \qquad
\al^{*}_{i}\rightarrow \ {\rm conjugates \ to} \ da_{i}.
\end{gather*}
In order for the correspondence be well def\/ined we need to
specify how the f\/ields (\ref{2.btgm}) change under a change of
coordinates $a_{1},\dots,a_{I}$. For the new set of coordinates
\begin{gather*}
\tld{a}_{i}=g_{i}(a_{1},\dots,a_{I}), \qquad
a_{i}=f_{i}(\tld{a}_{1},\dots,\tld{a}_{I})
\end{gather*}
this is determined in~\cite{MSV} by the formulas
\begin{gather}
\tld{a}_{i}(z)=g_{i}(a_{1}(z),\dots ,a_{I}(z)),
\nonumber\\
\tld{\al}_{i}(z)=g_{i,j}(a_{1}(z),\dots ,a_{I}(z))\al_{j}(z), \
\tld{\al}^{*}_{i}(z)=f_{i,j}(a_{1}(z),\dots
,a_{I}(z))\al^{*}_{j}(z),\label{2.btgmchng}
\\
\tld{a}^{*}_{i}(z)=f_{i,j}(a_{1}(z),{\dots},a_{I}(z))a^{*}_{j}(z){+}
f_{n,i,l}(a_{1}(z),{\dots},a_{I}(z))g_{l,m}(a_{1}(z),{\dots},a_{I}(z))\al^{*}_{n}(z)\al_{m}(z),
\nonumber
\end{gather}
where $g_{i,j}={\d g_{i}\ov \d a_{j}}$, $f_{i,j}={\d f_{i}\ov \d
\tld{a}_{j}}$, $f_{i,j,k}={\d^{2} f_{i}\ov \d \tld{a}_{j}\d
\tld{a}_{k}}$ and the normal ordering of the operators is implied.

Then the $N=2$ Virasoro superalgebra takes the form
\begin{gather*}
G^{-}=\sum_{i}\al_{i}a^{*}_{i}, \qquad
G^{+}=\sum_{i}\left(1-\tfrac{1}{\mu_{i}}\right)\al^{*}_{i}\d
a_{i}-\tfrac{1}{\mu_{i}}a_{i}\d\al^{*}_{i}, \nonumber\\
J=\sum_{i}\left(1-\tfrac{1}{\mu_{i}}\right)\al^{*}_{i}\al_{i}+\tfrac{1}{\mu_{i}}a_{i}a^{*}_{i},\nonumber\\
T=\sum_{i}\tfrac{1}{2}\left(\left(1+\tfrac{1}{\mu_{i}}\right)\d\al^{*}_{i}\al_{i}-
\left(1-\tfrac{1}{\mu_{i}}\right)\al^{*}_{i}\d\al_{i}\right)+
\left(1-\tfrac{1}{2\mu_{i}}\right)\d a_{i}a^{*}_{i}-
\tfrac{1}{2\mu_{i}}a_{i}\d a^{*}_{i}.
\end{gather*}
We see that $G^{-}(z)$ is the string version of de Rham
dif\/ferential on ${\mathbb C}^{I}$ and due to (\ref{2.btgmchng})
we can see the space of states generated by the f\/ields
(\ref{2.btgm}) as a chiral de Rham complex on ${\mathbb C}^{I}$
introduced recently by Malikov, Shechtman and Vaintrob~\cite{MSV}
for smooth manifolds. It is easy to see from (\ref{2.amp}) that
GSO projection in the open string sector makes the orbifold of
${\mathbb C}^{I}$. Indeed, the operator $\exp{(2\im\pi mJ[0])}$ in
the summation over $m$ produces the projection of the space of
states generated by $bc\bet\gm$-f\/ields on the subspace with
integer $J[0]$-charge, while the operator $U^{n\bv}$ in the
summation over $n$ generates the twisted sectors for the
$bc\bet\gm$-f\/ields. It allows to conclude that open string
sector can be described in terms of chiral de Rham complex on the
orbifold ${\mathbb C}^{I}/{\rm GSO}$. This object has been
introduced recently by Frenkel and Szczesny in~\cite{FSch}.

 Now we take the second step in the cohomology calculation, i.e.\
we calculate the cohomology with respect to
$\sum_{i}Q^{-}_{i}$-dif\/ferential. The screening charges
$Q^{-}_{i}$ can be expressed in terms of the f\/ields
(\ref{2.btgm}) as
\begin{gather*}
Q^{-}_{i}=\oint dz \al_{i}a^{\mu_{i}-1}_{i}(z).
\end{gather*}
Therefore $\sum_{i}Q^{-}_{i}$ is Koszul dif\/ferential associated
with Landau--Ginzburg potential
\begin{gather*}
W(a_{i})=\sum_{i}a^{\mu_{i}}_{i}
\end{gather*}
It proves that $A$-type D-branes in Gepner model are fractional
D-branes~\cite{DGom} on the Landau--Ginzburg orbifold. Notice
that taking the $\sum_{i}Q^{-}_{i}$ cohomology as the f\/irst step
we get the same background geometry of ${\mathbb C}^{I}/GSO$
Landau--Ginzburg orbifold~\cite{SP2}.
The similar analysis can be carried out also for the case of
B-type branes~\cite{SP2}.

Thus, a free f\/ield construction of D-branes allows to extract
geometry from the purely algebraic Recknagel--Schomerus
construction. It is interesting to develop in more details the
D-brane geometry description in terms of chiral de Rham complex.

\subsection*{Acknowledgements}

I thank the organizers of the Symmetry-2007 conference for their
hospitality. This work was supported in part by grants
RFBR-07-02-0799-a, SS6358.2006.2.

\pdfbookmark[1]{References}{ref}
\LastPageEnding

\end{document}